\documentclass[pre,twocolumn,showpacs]{revtex4}
\usepackage{graphicx}
\usepackage{amsmath,amssymb}
\usepackage{color}

\newcommand{\myav}[1]{\langle #1\rangle}

\newcommand{\latin}[1]{{\itshape #1}}

\newcommand{\eg}{\latin{e.\,g.}}

\newcommand{\ie}{\latin{i.\,e.}}

\newcommand{\inextremis}{\latin{in extremis}}
\newcommand{\etal}{\latin{et al.}}

\begin{document}

\title{Biodiversity on island chains: neutral model simulations}

\author{Patrick B. Warren}
\affiliation{Unilever R\&D Port Sunlight, Bebington, Wirral, CH63 3JW, UK.}

\date{April 6, 2010}

\begin{abstract}
A neutral ecology model is simulated on an island chain, in which
neighbouring islands can exchange individuals but only the first
island is able to receive immigrants from a metacommunity.  It is
found by several measures that biodiversity decreases along the chain,
being highest for the first island.  Subtle changes in taxon abundance
distributions can be detected when islands in the chain are compared
to diversity-matched single islands.  The results potentially apply to
human microbial diversity, but highlight the difficulty of using
static single-site taxon abundance distributions to discriminate
between dispersal limitation mechanisms.
\end{abstract}

\pacs{87.23.-n, 87.10.Mn, 02.50.Ga}


\maketitle

It has recently been observed that human microbial diversity varies
systematically between body sites \cite{XGKC+09, XCLH+09}, for example
phylogenetic diversity is higher for the palm of the hand and the sole
of the foot, than for the armpit and forehead (Fig.~S14 in
Ref.~\onlinecite{XCLH+09}).  A high degree of inter-individual
variability is also observed, to the point where the characterisation
of residual skin bacteria has been proposed as a novel forensic tool
\cite{XFLZ+10}.  The latter, in particular, supports the notion that
\emph{stochastic dispersal limitation} may play a significant role in
determining microbial diversity.  Stochastic dispersal limitation is a
signature element of Hubbell's unified neutral model of biodiversity
and biogeography \cite{Hub01}, and this motivates the question of
whether neutral models can be applied to human microbial biodiversity
and biogeography.  This is a hard problem and I do not claim to have
solved it here.  Rather, the present study is restricted to exploring
the role of dispersal mechanisms in the context of neutral theory,
keeping in mind the possible application to the human microbiome.

The merits of neutral models have been debated extensively
\cite{Bel01-Lei07}, and it is far from obvious that they should apply
to human microbiota \cite{XFGM+08, niche}.  However Hubbell's neutral
model has recently been successfully applied to predict microbial
diversity in tree holes \cite{XBAS+05, XWvdG+07}.  In this context it
is important to note that it has been argued dispersal limitation is
the dominant factor determining taxon abundancies \cite{EA05, EAM07},
with other neutral model ideas, such as the zero-sum constraint
(single trophic level; community saturation) or speciation by point
mutation, playing a lesser role.

\begin{figure}
\begin{center}
\includegraphics{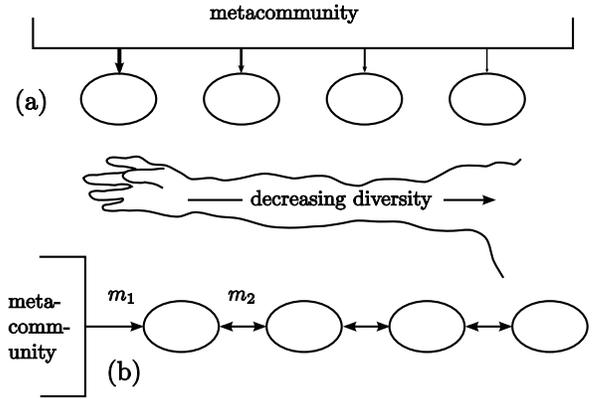}
\end{center}
\caption[?]{Putative explanations for a variation in human microbial
  diversity, based on dispersal-limitation and the theory of island
  biogeography: (a) variable-immigration-rate hypothesis and (b)
  island-chain hypothesis.\label{fig:hypo}}
\end{figure}

If taxon abundancies are largely determined by stochastic dispersal
limitation, then a couple of limiting hypotheses (Fig.~\ref{fig:hypo})
present themselves to explain the observed variations in human
microbial diversity.  The first is a \emph{variable-immigration-rate
  hypothesis} in which different body sites are envisaged as being
microbial `islands' in contact with a microbial metacommunity but
effectively isolated from each other.  Here variation in diversity
corresponds to a variable immigration rate from the metacommunity.
The second hypothesis is an \emph{island-chain hypothesis} in which it
is envisaged that migration can take place between islands but,
\inextremis, it is only the first island (\eg\ the hand) that receives
immigrants from the metacommunity.  In this case one expects that
diversity should decrease as one moves along the chain away from the
island in contact with the metacommunity, due to dispersal limitation.
This is confirmed in the present study.

Of course these hypotheses represent limiting cases and, if dispersal
limitation is relevant, reality probably lies somewhere in between.  A
second question therefore is whether one can use taxon abundance
distributions to distinguish between dispersal mechanisms.
Unfortunately, the present study finds that both hypotheses lead to
rather similar abundance distributions.  When this is conflated with
other factors, such as deviations from neutral model dynamics
\cite{BFF09}, it is probably going to be difficult to distinguish
between dispersal mechanisms on the basis of static single-site
measurements of microbial diversity.

The neutral model has been extensively studied for isolated islands in
contact with a metacommunity \cite{Hub01, VH03, XVBH+03, MAS04, EA05,
  He05, EA06, VH06, Bab06}, but only for certain cases has it been
solved for multiple islands, or `patches', which are able to exchange
individuals \cite{CL02, XCPL+02}.  In particular, the island chain
problem has not been solved (\ie\ where individuals can migrate
between neighbouring islands but immigration is restricted to the
first island in the chain).  The primary goal of the present study is
to solve this problem.  Although in principle one can approach this
analytically, the experience of Vallade and Houchmandzadeh \cite{VH06}
for \emph{two} islands suggests this will be effectively unmanageable.
I therefore approach the problem by means of simulations.

Let me start by summarising the mathematical characterisation of taxon
abundance distributions.  Suppose there are $K$ taxa and $N_i$
individuals in the $i$-th taxon ($i=1\dots K$), in a population of
$J=\sum_{i=1}^K N_i$ individuals.  The relative abundance of the
$i$-th taxon is defined to be $\omega_i = N_i/J$.  The taxon abundance
distribution is characterised by $\phi_k$, the number of taxa
containing $k$ individuals.  Formally $\phi_k=\sum_{i=1}^K \delta_{k,
  N_i}$ where $\delta_{n, m}$ takes the value unity if $n = m$ and is
zero otherwise.  Given the set of $N_i$ one can easily calculate
$\phi_k$.  One has $K = \sum_{k=1}^\infty \phi_k$ and $J =
\sum_{k=1}^\infty k \phi_k$.  Since no taxon can contain more
individuals than there are in the community as a whole, $\phi_k=0$ for
$k>J$. Similarly $\phi_J=1$ if and only if all the individuals belong
to the same taxon (the `monodominated' state), otherwise $\phi_J=0$.

In standard neutral model dynamics, population sizes remain fixed
(saturated) and are specified at the outset, whilst the number of taxa
and the number of individuals per taxon fluctuates.  I adopt the
notation of Vallade and Houchmandzadeh \cite{VH03, VH06} and write
$\myav{\cdots}$ to indicate a quantity averaged over an ensemble of
populations undergoing neutral model dynamics.  The information in
$\myav{\phi_k}$ is conveniently represented by giving the
ensemble-average probability $p(\omega)$ that an individual belongs to
a taxon of relative abundance $\omega$ \cite{VH03, VH06, XTip79}.  For
a community of a finite size, $p(\omega)$ is a discrete array or
`comb' of $\delta$-functions, even after ensemble-averaging, since
$\omega$ can only take on discrete multiples of $1/J$.  However as
$J\to\infty$, $p(\omega)$ becomes a continuous function.  One can show
that the continuum limit is $p(\omega)=\lim_{J\to\infty}
k\myav{\phi_{k}}$ where $k=\omega J$ \cite{VH03}.

I shall additionally use several ensemble-average measures of
diversity.  The principal one of these is the Simpson diversity index
\cite{Sim49}, defined for a given set of taxon abundancies to be $D =
1-\sum_{i=1}^K \omega_i^2$.  It is related to the second moment of the
taxon abundance distribution by $D=1-J^{-2}\sum_{k=1}^\infty
k^2\phi_k$. From this it can easily be shown that in the continuum
limit
\begin{equation}
\textstyle \myav{D}=1- \int_0^1\omega\,p(\omega)\,d\omega.\label{eq:ddef}
\end{equation}
The second diversity measure is the ensemble-average number of
taxa $\myav{K}=\sum_{k=1}^J\myav{\phi_k}$.  The third is the
ensemble-average monodominance probability $\myav{\phi_J}$---as
explained above $\phi_J$ is 1 or 0 according to whether or not all the
individuals belong to the same taxon.

As an order parameter, the Simpson index $\myav{D}$ has some
advantages over $\myav{K}$ and $\myav{\phi_J}$: it remains well
defined in the continuum limit $J\to\infty$, there are some
particularly simple theoretical expressions for $\myav{D}$ under
neutral model dynamics, and in particular there is a prediction
(confirmed by simulation) that $\myav{D}$ factorises into a product of
the metacommunity diversity index, and an island factor.  Also the
Simpson diversity index generalises naturally to a measure of
$\beta$-diversity \cite{CL02}, and to time-series data \cite{XAPB+06}.
The index satisfies $0\le\myav{D}\le1-1/K$.  There is a mild
disadvantage in that $\myav{D}$ loses sensitivity to the underlying
abundance distribution at the limiting values.

\begin{figure}
\begin{center}
\includegraphics{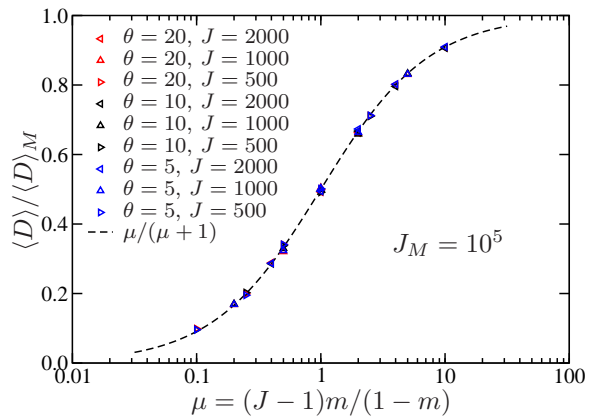}
\end{center}
\caption[?]{Steady state island diversity $\myav{D}$ from neutral
  model simulations on islands of various sizes with varying
  immigration rates, normalised by the metacommunity diversity
  $\myav{D}_M=\theta/(\theta+1)$.  The agreement with the theory,
  Eq.~\eqref{eq:idiv}, is excellent.  Error bars are smaller than the
  symbols.\label{fig:idiv}}
\end{figure}

Let me next summarise neutral model dynamics.  In the metacommunity
it is as follows.  An individual is selected at random,
and with probability $1-\nu$ is replaced with a copy of another
individual drawn at random from the metacommunity, or with probability
$\nu$ is replaced by an individual belonging to a new taxon.  Thus
$\nu$ is the speciation rate.  For $\nu=0$ the metacommunity
eventually falls into a monodominated state, in an ecological analog
of the Matthew principle \cite{matthew}.  For $\nu>0$ the taxon
abundance distribution is a balance between speciation and extinction.

An explicit expression for the taxon abundance distribution in a
metacommunity of size $J_M$ has been obtained by a number of workers
\cite{Hub01, VH03, MAS04, He05, EA05, EA06}.  Results are
quoted as \emph{metacommunity} (subscript `M') steady-state
ensemble-averages:
\begin{equation}
\myav{\phi_k}_M=\frac{\theta\,\Gamma(J_M+1)\,\Gamma(J_M+\theta-k)}%
{k\,\Gamma(J_M+1-k)\,\Gamma(J_M+\theta)}\label{eq:mphik}
\end{equation}
where $\theta=(J_M-1)\nu/(1-\nu)$.  One has $\theta \approx J_M\nu$
for $J_M\gg1$ and $\nu\ll1$.  It can be shown that
$\myav{J}_M=\sum_{k=1}^{J_M}k\myav{\phi_k}_M=J_M$ (an identity), and
$\myav{K}_M = \sum_{k=1}^{J_M}\myav{\phi_k}_M =
\sum_{k=1}^{J_M}\frac{\theta}{\theta-1+k}$.  The continuum limit of
Eq.~\eqref{eq:mphik} can be obtained using Stirling's approximation.
One finds $p(\omega)=\theta(1-\omega)^{\theta-1}$.  It follows that
the metacommunity diversity order parameter in the continuum limit is
\begin{equation}
\myav{D}_M = \frac{\theta}{\theta+1}.\label{eq:md}
\end{equation}
This result was noted by He and Hu by analogy to a similar problem in
genetics \cite{HH05}.  

Neutral model dynamics on an island connected to the metacommunity is
as follows.  An individual is selected at random, and with probability
$1-m$ is replaced with a copy of another individual drawn at random
from the island, or with probability $m$ is replaced by an individual
drawn at random from the metacommunity.  Thus $m$ is the immigration
rate.  Similar to the metacommunity, the island community
eventually falls into a monodominated state if $m=0$, whereas for $m>0$ a
steady-state taxon abundance distribution arises as a balance between
immigration and extinction.  It is often a very good approximation to
assume that the island dynamics are decoupled from the metacommunity
dynamics; in other words the metacommunity can be taken to
have a static abundance distribution.  This is because the
metacommunity abundance distribution turns over on a time scale of
order $1/\nu$ whereas the island abundance distribution turns over
on a time scale of order $1/m$, and typically $\nu\ll m$.  Here time
scales are quoted in terms of the number of replacement steps per
individual, since this is expected to be proportional to the real
elapsed time \cite{Hub01}.

\begin{figure}
\begin{center}
\includegraphics{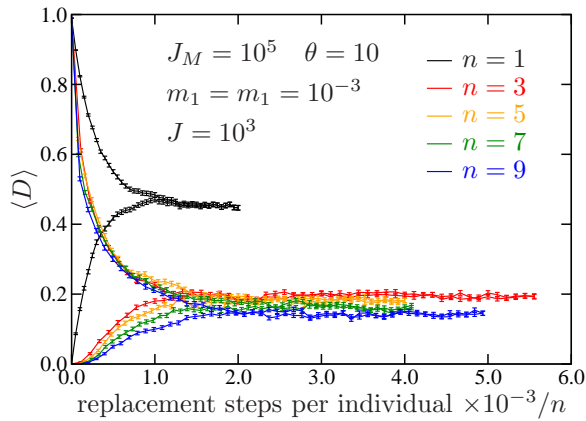}
\end{center}
\caption[?]{Approach to steady state for the central island of a chain
  of length $n$ islands starting from either a completely monodominated
  state (lower curves), or a flat abundance distribution (upper curves;
  flat means the $N_i$ are equalised subject to $\sum_{i=1}^KN_i=J$).
  The $n=1$ case is for a single island.  For $n>1$ the steady state
  is weakly dependent on $n$.\label{fig:atc}}
\end{figure}

Exact results for the island taxon abundance distribution were
obtained only recently \cite{EA05, EA06}, although partial results
were obtained by previous authors \cite{XVBH+03, VH03, MAS04}.  The
result is
\begin{equation}
\myav{\phi_k}=\binom{J}{k}\int_0^1 \!\frac{du}{u}\theta(1-u)^{\theta-1}
\frac{(\mu u)_k(\mu(1-u))_{J-k}}{(\mu)_J}
\label{eq:iphik}
\end{equation}
where $(x)_n=\Gamma(x+n)/\Gamma(x)$ is the Pochhammer symbol,
$\binom{J}{k} = \Gamma(J+1)/(\Gamma(k+1)\,\Gamma(J-k+1))$ is the
binomial coefficient, and $\mu=m(J-1)/(1-m)$ plays a role similar to
$\theta$ for the metacommunity.  For $J\gg1$ and $m\ll1$, one has
$\mu=Jm$.  Note that $J_M$ does not feature in this expression, in
other words the island abundance distribution is insensible to the
metacommunity size.  This point is discussed more thoroughly by
Vallade and Houchmandzadeh \cite{VH06}.  Eq.~\eqref{eq:iphik}
simplifies in the limit $k=J$ to give an expression for the island
monodominance probability,
\begin{equation}
\myav{\phi_J}=\int_0^1 \!\frac{du}{u}\theta(1-u)^{\theta-1}
\frac{(\mu u)_J}{(\mu)_J}.\label{eq:iphiJ}
\end{equation}
This depends strongly on all the relevant parameters and vanishes
asymptotically for $J\to\infty$ at fixed $\mu$ as $\myav{\phi_J}\sim
\theta\Gamma(\theta)(\mu\ln J)^{-\theta}$.  The
continuum limit of Eq.~\eqref{eq:iphik} is \cite{VH03, AM04, EA05,
  EA06}
\begin{equation}
p(\omega)=\mu\theta\int_0^1 
\binom{\mu}{\mu u}
\,(1-\omega)^{\mu u-1}
\omega^{\mu(1-u)} u^{\theta} du\,.\label{eq:ip}
\end{equation}
Inserting
Eq.~\eqref{eq:ip} into Eq.~\eqref{eq:ddef} gives a simple but to my
knowledge previously unreported result,
\begin{equation}
\myav{D}=\frac{\mu\theta}{(\mu+1)(\theta+1)}.\label{eq:idiv}
\end{equation}
Remarkably, as alluded to above, the diversity index factorises into
the product of the metacommunity diversity index
$\myav{D}_M=\theta/(\theta+1)$ and an island factor
$\mu/(\mu+1)$.

\begin{figure}
\begin{center}
\includegraphics{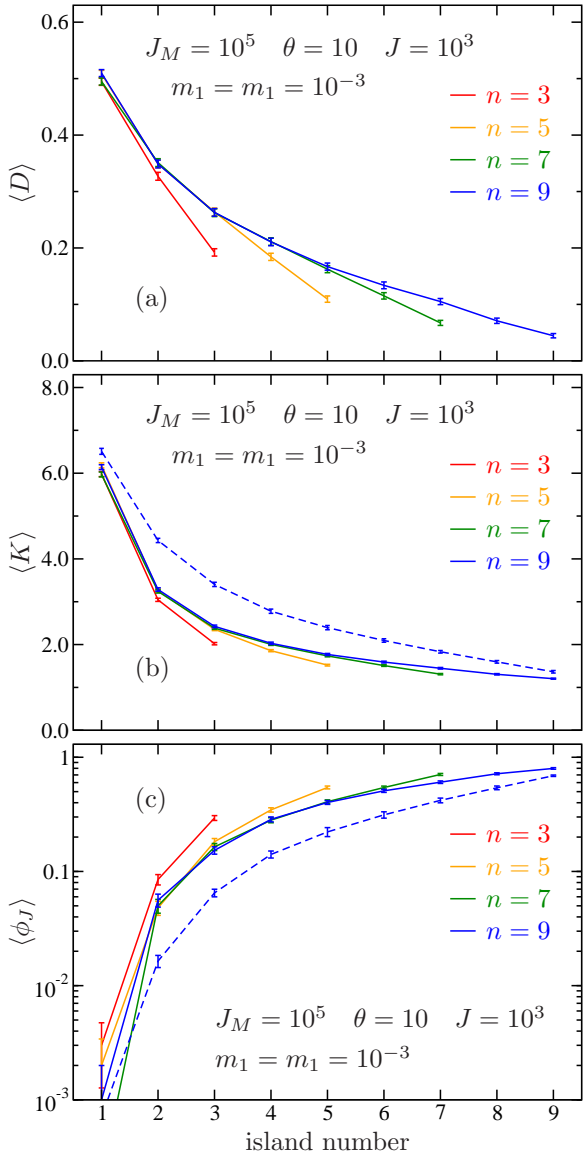}
\end{center}
\caption[?]{Aspects of diversity on island chains of length $n$: (a)
  diversity index $\myav{D}$, (b) number of taxa $\myav{K}$, and (c)
  monodominance probability $\myav{\phi_J}$.  The dashed lines in (b)
  and (c) are the $\myav{D}$-matched single island results for $n=9$
  (see text).\label{fig:dkp}}
\end{figure}

\begin{figure}
\begin{center}
\includegraphics{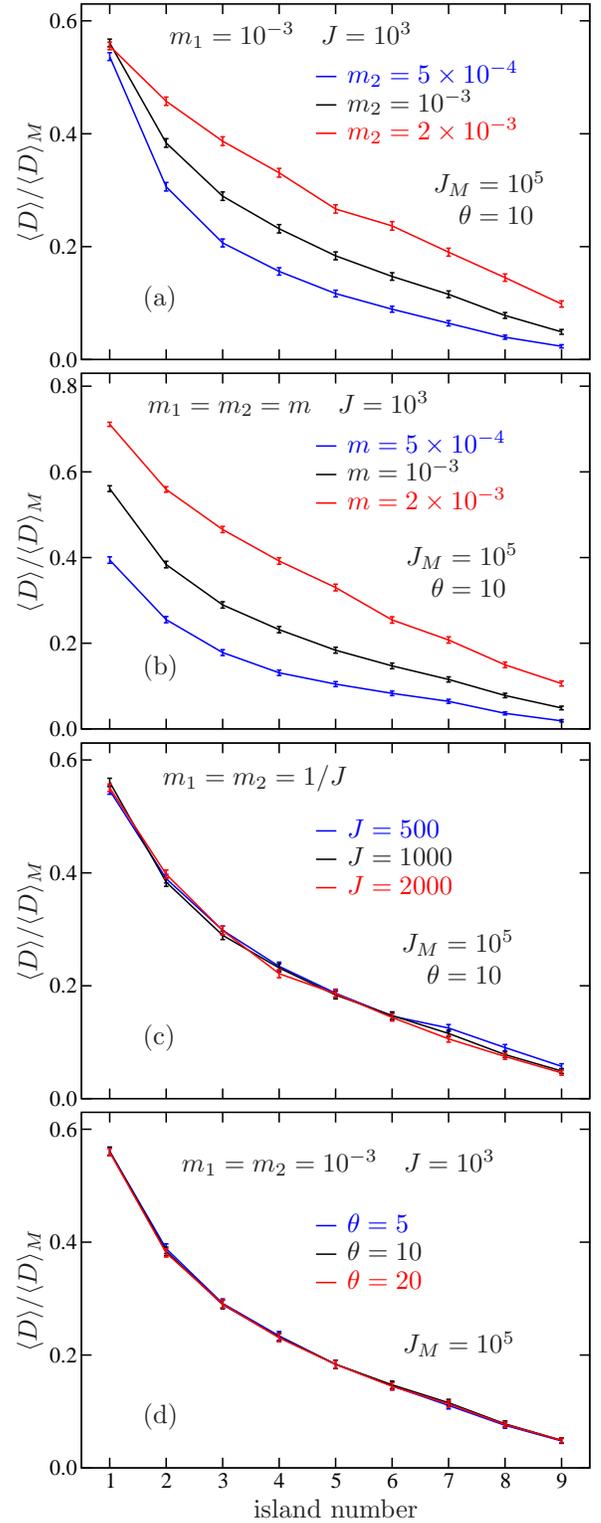}
\end{center}
\caption[?]{Diversity index $\myav{D}$, normalised by the theoretical
  metacommunity diversity index $\myav{D}_M=\theta/(\theta+1)$, along
  a chain of $n=9$ islands: (a) varying inter-island migration rate
  $m_2$ only, (b) varying migration and immigration rates together at
  fixed island size, (c) varying migration and immigration rates
  inversely with island size, (d) varying metacommunity diversity
  parameter.\label{fig:var}}
\end{figure}

Simulation of neutral model dynamics as summarised above is
straightforward.  I make the assumption that metacommunity and island
dynamics are decoupled (discussed in more detail below).  Therefore I
generate a large number ($10^3$--$10^5$) of metacommunity abundance
distribution samples for given $J_M$ and $\theta$, equilibrating for
$10\times J_M^2/\theta$ replacement steps between samples to ensure
statistical independence \cite{VH06}.  I use these samples in
subsequent island and island chain simulations.  As a reference point,
I shall use $\theta=10$, motivated by Woodcock \etal\ \cite{XWvdG+07},
and $J_M=10^5$, motivated not so much by time scale considerations
(see later) but by the requirement that $J_M\gg J\gg1$ \cite{VH06}.
Except where otherwise stated, averages are over $10^3$ samples.

I undertook a number of single island simulations to build confidence
in the simulation and analysis methodologies.  I find excellent
agreement between these simulations and the theoretical predictions
for the steady-state properties (equilibrating for $10\times J/m$
replacement steps between samples).  For example Fig.~\ref{fig:idiv}
compares theory and simulation results for the Simpson diversity
index.

\begin{figure}
\begin{center}
\includegraphics{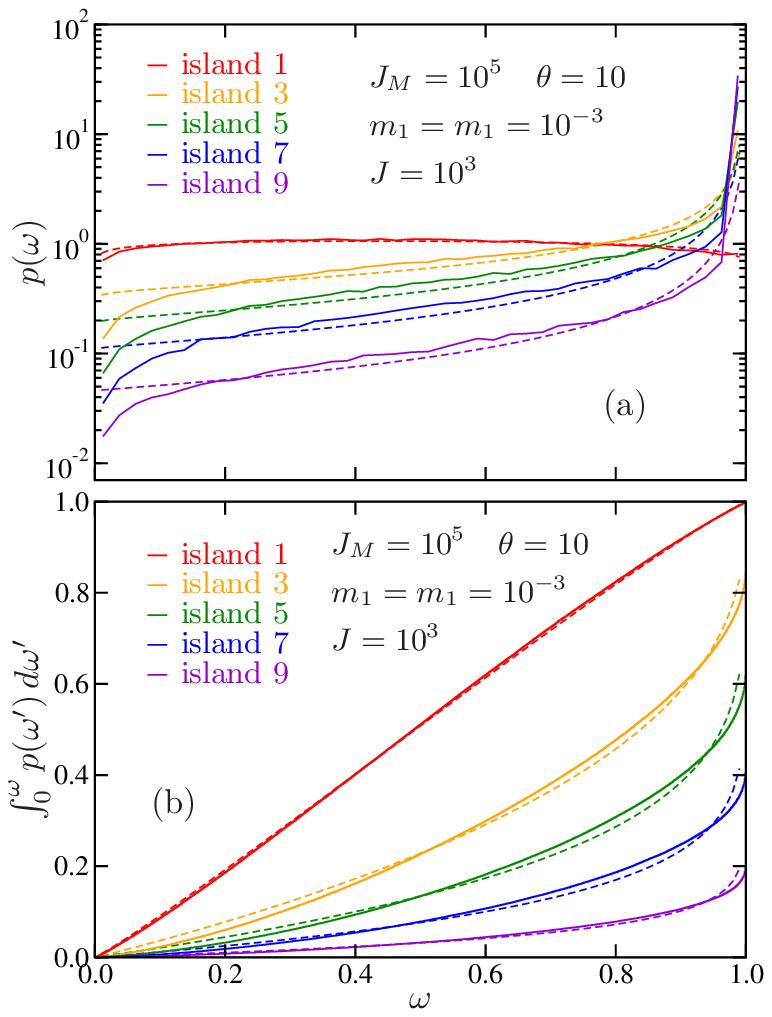}
\end{center}
\caption[?]{Abundance distributions for selected islands in an island
  chain of length $n=9$: (a) probability that a randomly selected
  individual belongs to a taxon of relative abundance $\omega$, and
  (b) cumulative distribution function of the same.  The results
  (solid lines) are compared to the theoretical expectations for
  $\myav{D}$-matched single islands calculated from Eq.~\eqref{eq:ip}
  (dashed lines).  Results
  are averages over $10^5$ samples.\label{fig:tad}}
\end{figure}

The island chain simulations are performed similarly to the single
island simulations.  I introduce an immigration rate $m_1$ (for the
first island) and an inter-island migration rate $m_2$.  This is
illustrated in Fig.~\ref{fig:hypo}(b).  Specifically, the dynamics are
as follows.  An individual is selected at random.  If the chosen
individual lies on the first island, it is replaced with a copy of
another individual on the island with probability $1-m_1-m_2$, an
immigrant from the metacommunity with probability $m_1$, or a migrant
from the neighbouring island with probability $m_2$.  If the chosen
individual lies on an island interior to the chain, it is replaced
with a copy of another individual on the island with probability
$1-2m_2$, or with a migrant from one of the neighbouring islands
(selected at random) with probability $2m_2$. If the chosen individual
lies on the terminal island, it is replaced with a copy of another
individual on the island with probability $1-m_2$, or with a migrant
from neighbouring island with probability $m_2$.  Migrants are copies
of individuals chosen at random on neighbouring islands.

Fig.~\ref{fig:atc} shows that the approach to steady state of a chain
of $n$ islands is slowed by a factor $\approx 1/n$ compared to the
single island case.  I therefore equilibrate the island chains against
each metacommunity sample for at least $10\times n^2J/m$ replacement
steps between samples, where $m$ is the smaller of $m_1$ and $m_2$.
In terms of the number of replacement steps per individual, the island
chain relaxation time scale is of the order $n/m$.  Clearly for
$n\gtrsim10$ the assumed time scale separation between this and the
metacommunity relaxation time $1/\nu\approx J_M/\theta\approx10^{4}$
is faltering.  Nevertheless the results are still valid provided they
can be shown to be unaffected by varying $J_M$ since it can be assumed
that in reality $J_M$ is much larger than $10^5$ \cite{XWvdG+07}.  To
test this, I repeated many of the simulations with $J_M=5\times10^4$
and $J_M=2\times10^5$.  I found this made no difference to the
measured island properties, within the statistical errors.

For a chain of length $n = 9$, I examined in some more detail how the
islands recover their steady state diversity.  The picture is a little
mixed.  The curves can be fitted by $\myav{D} = A + B e^{-t/\tau}$,
but not very well, indicating there is not a clearly dominant
relaxation time.  Moreover the fitted value of $\tau$ is affected by
whether one starts with a monodominated state or a uniform state. What
this all suggests is that there is a spectrum of relaxation modes,
which are excited differently according to the initialisation
protocol, and which are subsequently mixed up by the non-linear
dynamics.  A more detailed exploration of this is left for future
work.

Representative steady-state results for the island chain simulations
are shown in Figs.~\ref{fig:dkp}--\ref{fig:tad}.  The first conclusion
(Fig.~\ref{fig:dkp}) is that diversity decreases, by whatever measure,
as one moves away from the island in contact with the metacommunity.
Fig.~\ref{fig:var} shows how island diversity varies with immigration
and migration rates $m_1$ and $m_2$, island size $J$, and the value of
$\theta$.  Increasing the inter-island migration rate $m_2$
(Fig.~\ref{fig:var}(a)) has the effect of increasing the diversity
along the island chain apart from the first island.  Additionally
increasing the metacommunity immigration rate $m_1$
(Fig.~\ref{fig:var}(b)) leads to increased diversity along the whole
chain.  Fig.~\ref{fig:var}(c) supports the notion that the island
diversity is governed by the combinations $Jm_1$ and $Jm_2$ rather
than the individual values of $J$, $m_1$ and $m_2$, in close analogy
to the theory for the single island.  Similarly Fig.~\ref{fig:var}(d)
strongly suggests that the Simpson diversity index continues to be
factorisable into the metacommunity diversity index multiplied by a
contribution from the structure of the island chain, again in close
analogy to the single island result.

I next compare islands in the chain to `$\myav{D}$-matched' single
islands.  Here $\myav{D}$-matching means a value for $\mu$ is inferred
from Eq.~\eqref{eq:idiv} (\ie\ $\mu=\myav{D}/(\myav{D}_M-\myav{D})$
where $\myav{D}_M=\theta/(\theta+1)$), and used to calculate values of
$\myav{K}=\sum_{k=1}^J\myav{\phi_k}$ and $\myav{\phi_J}$ from
Eqs.~\eqref{eq:iphik} and \eqref{eq:iphiJ}.  I assume the island size
$J$ is fixed.  The procedure amounts to matching the first and second
moments of $\myav{\phi_k}$.  The dashed lines in Fig.~\ref{fig:dkp}(b)
and (c) show systematically that the ensemble-average number of taxa
is reduced and the monodominance probability is increased, comparing
an island in the island chain with its $\myav{D}$-matched single
island counterpart.  Thus there is a tendency towards \emph{fewer,
  larger taxa}, when islands in a chain are compared to
$\myav{D}$-matched single islands.

A more detailed examination of the abundance distributions shows that
there is a subtle and non-trivial redistribution of the taxon
abundancies.  When compared to the $\myav{D}$-matched single islands,
Fig.~\ref{fig:tad} shows that $p(\omega)$ is reduced for
$\omega\lesssim0.2$ and $\omega\gtrsim0.8$, but increased for
$0.2\lesssim\omega\lesssim0.8$.  This means that the number of taxa
with intermediate abundancies is increased at the expense of the very
rare taxa and the high abundance taxa.  But, in addition, the
cumulative distribution function jumps up at $\omega=1$, as shown
clearly in Fig.~\ref{fig:tad}(b).  This corresponds to the increased
monodominance probability.  At first sight this is at odds with with
the redistribution towards mid-range abundancies, nevertheless it is a
real effect and indeed is the reason why monodomination was separately
studied.

The loss of the very rare taxa can perhaps be attributed to the
filtering properties of the island chain.  These taxa are already rare
in the metacommunity and it could simply be that a representative from
a rare taxon is less likely to arrive via migration along an island
chain than via direct immigration from the metacommunity (at matched
$\myav{D}$).  The loss of the high abundance taxa and the increased
monodominance probability are more mysterious and I do not at present
have a clear mechanistic explanation.  Possibly what is happening for
islands in a chain, compared to $\myav{D}$-matched single islands, is
that the monodominated state ($\omega=1$) has become `stickier' in
dynamical terms, without actually becoming an adsorbing state.  In the
monodominated state there is of course only one taxon, with
$\omega=1$, and this may come at the expense of the high abundance
taxa with $0.8\lesssim\omega<1$.

For $\myav{D}\lesssim0.5$ the abundance distribution
for a $\myav{D}$-matched \emph{metacommunity} is almost exactly the
same as that for a $\myav{D}$-matched single island.  By this I mean
that $p(\omega)=\theta(1-\omega)^{\theta-1}$ with
$\theta=\myav{D}/(1-\myav{D})$ is a very good approximation to
$p(\omega)$ from Eq.~\eqref{eq:ip}.  However a complete
comparison with an equivalent metacommunity is frustrated by the
residual dependence of $\myav{K}_M$ and $\myav{\phi_J}_M$ on the
metacommunity size $J_M$.

Despite these subtleties, it is clear from Fig.~\ref{fig:tad} that the
taxon abundance distributions on an island chain are quite well
approximated by $\myav{D}$-matched single islands.  This is the origin
of the claim in the introduction that it is probably going to be
difficult to distinguish between dispersal mechanisms on the basis of
static single-site measurements of the taxon abundancies.  To resolve
this question, or indeed to distinguish between dispersal-limitation
and niche-adaptation \cite{niche}, probably requires more detailed
examination of the $\beta$-diversity \cite{CL02, XCPL+02, DCH06}, and
dynamics \cite{XAPB+06, Bab06, VH06}.  In this context it may be
useful to explore spatial correlations \cite{CL02} and temporal
correlations \cite{XAPB+06}, which are natural generalisations of the
Simpson index $D$.

I thank Mike Cates, Chris Quince, Bill Sloan, and Dave Taylor, for
helpful comments.


\begin{thebibliography}{28}
\expandafter\ifx\csname natexlab\endcsname\relax\def\natexlab#1{#1}\fi
\expandafter\ifx\csname bibnamefont\endcsname\relax
  \def\bibnamefont#1{#1}\fi
\expandafter\ifx\csname bibfnamefont\endcsname\relax
  \def\bibfnamefont#1{#1}\fi
\expandafter\ifx\csname citenamefont\endcsname\relax
  \def\citenamefont#1{#1}\fi
\expandafter\ifx\csname url\endcsname\relax
  \def\url#1{\texttt{#1}}\fi
\expandafter\ifx\csname urlprefix\endcsname\relax\def\urlprefix{URL }\fi
\providecommand{\bibinfo}[2]{#2}
\providecommand{\eprint}[2][]{\url{#2}}

\bibitem[{XGK()}]{XGKC+09}
\bibinfo{note}{\bibinfo{author}{\bibfnamefont{E.~A.} \bibnamefont{Grice}},
  \etal, \bibinfo{journal}{Science} \textbf{\bibinfo{volume}{324}},
  \bibinfo{pages}{1190} (\bibinfo{year}{2009}).}

\bibitem[{XCL()}]{XCLH+09}
\bibinfo{note}{\bibinfo{author}{\bibfnamefont{E.~K.} \bibnamefont{Costello}},
  \etal, \bibinfo{journal}{Science} \textbf{\bibinfo{volume}{326}},
  \bibinfo{pages}{1694} (\bibinfo{year}{2009}).}

\bibitem[{XFL()}]{XFLZ+10}
\bibinfo{note}{\bibinfo{author}{\bibfnamefont{N.}~\bibnamefont{Fierer}}, \etal,
  \bibinfo{journal}{Proc. Natl. Acad. Sci. (USA)} (\bibinfo{year}{2010}),
  \bibinfo{note}{doi:10.1073/pnas.1000162107}.}

\bibitem[{\citenamefont{Hubbell}(2001)}]{Hub01}
\bibinfo{author}{\bibfnamefont{S.~P.} \bibnamefont{Hubbell}},
  \emph{\bibinfo{title}{The unified neutral theory of biodiversity and
  biogeography}} (\bibinfo{publisher}{Princeton University press},
  \bibinfo{address}{Princeton, NJ}, \bibinfo{year}{2001}).

\bibitem[{Bel()}]{Bel01-Lei07}
\bibinfo{note}{\bibinfo{author}{\bibfnamefont{G.}~\bibnamefont{Bell}},
  \bibinfo{journal}{Science} \textbf{\bibinfo{volume}{293}},
  \bibinfo{pages}{2413} (\bibinfo{year}{2001});
  \bibinfo{author}{\bibfnamefont{B.~J.} \bibnamefont{McGill}},
  \bibinfo{journal}{Nature} \textbf{\bibinfo{volume}{422}},
  \bibinfo{pages}{881} (\bibinfo{year}{2003});
  \bibinfo{author}{\bibfnamefont{J.}~\bibnamefont{Chave}},
  \bibinfo{journal}{Ecol. Lett.} \textbf{\bibinfo{volume}{7}},
  \bibinfo{pages}{241} (\bibinfo{year}{2004});
  \bibinfo{author}{\bibfnamefont{D.}~\bibnamefont{Alonso}},
  \bibinfo{author}{\bibfnamefont{R.~S.} \bibnamefont{Etienne}},
  \bibnamefont{and} \bibinfo{author}{\bibfnamefont{A.~J.}
  \bibnamefont{McKane}}, \bibinfo{journal}{TRENDS Ecol. Evol.}
  \textbf{\bibinfo{volume}{21}}, \bibinfo{pages}{451} (\bibinfo{year}{2006});
  \bibinfo{author}{\bibfnamefont{J.}~\bibnamefont{Chave}},
  \bibinfo{author}{\bibfnamefont{D.}~\bibnamefont{Alonso}}, \bibnamefont{and}
  \bibinfo{author}{\bibfnamefont{R.~S.} \bibnamefont{Etienne}},
  \bibinfo{journal}{Nature} \textbf{\bibinfo{volume}{441}}, \bibinfo{pages}{E1}
  (\bibinfo{year}{2006}); \bibinfo{author}{\bibfnamefont{E.~G.}
  \bibnamefont{Leigh}, \bibfnamefont{Jr.}}, \bibinfo{journal}{J. Evol. Biol.}
  \textbf{\bibinfo{volume}{20}}, \bibinfo{pages}{2075} (\bibinfo{year}{2007}).}

\bibitem[{XFG()}]{XFGM+08}
\bibinfo{note}{\bibinfo{author}{\bibfnamefont{B.}~\bibnamefont{Foxman}}, \etal,
  \bibinfo{journal}{Interdisc. Perspect. Infect. Diseases}
  (\bibinfo{year}{2008}), \bibinfo{note}{doi:10.1155/2008/613979}.}

\bibitem[{nic()}]{niche}
\bibinfo{note}{For human microbiota it seems quite likely that the most
  dominant taxa are niche-adapted, as for example {\it Staphylococcus
  epidermidis} on skin. However there is a large tail of rarer taxa for which
  dispersal limitation may be relevant.}

\bibitem[{XBA()}]{XBAS+05}
\bibinfo{note}{\bibinfo{author}{\bibfnamefont{T.}~\bibnamefont{Bell}}, \etal,
  \bibinfo{journal}{Science} \textbf{\bibinfo{volume}{308}},
  \bibinfo{pages}{1884} (\bibinfo{year}{2005}); see also
  \bibinfo{journal}{Science} \textbf{\bibinfo{volume}{309}},
  \bibinfo{pages}{Letters} (\bibinfo{year}{2005}).}

\bibitem[{XWv()}]{XWvdG+07}
\bibinfo{note}{\bibinfo{author}{\bibfnamefont{S.}~\bibnamefont{Woodcock}},
  \etal, \bibinfo{journal}{FEMS Microbiol. Ecol.}
  \textbf{\bibinfo{volume}{62}}, \bibinfo{pages}{171} (\bibinfo{year}{2007}).}

\bibitem[{\citenamefont{Etienne and Alonso}(2005)}]{EA05}
\bibinfo{author}{\bibfnamefont{R.~S.} \bibnamefont{Etienne}} \bibnamefont{and}
  \bibinfo{author}{\bibfnamefont{D.}~\bibnamefont{Alonso}},
  \bibinfo{journal}{Ecol. Lett.} \textbf{\bibinfo{volume}{8}},
  \bibinfo{pages}{1147} (\bibinfo{year}{2005}).

\bibitem[{\citenamefont{Etienne et~al.}(2007)\citenamefont{Etienne, Alonso, and
  McKane}}]{EAM07}
\bibinfo{author}{\bibfnamefont{R.~S.} \bibnamefont{Etienne}},
  \bibinfo{author}{\bibfnamefont{D.}~\bibnamefont{Alonso}}, \bibnamefont{and}
  \bibinfo{author}{\bibfnamefont{A.~J.} \bibnamefont{McKane}},
  \bibinfo{journal}{J. Theor. Biol.} \textbf{\bibinfo{volume}{248}},
  \bibinfo{pages}{522} (\bibinfo{year}{2007}).

\bibitem[{\citenamefont{Bianconi et~al.}(2009)\citenamefont{Bianconi, Ferretti,
  and Franz}}]{BFF09}
\bibinfo{author}{\bibfnamefont{G.}~\bibnamefont{Bianconi}},
  \bibinfo{author}{\bibfnamefont{L.}~\bibnamefont{Ferretti}}, \bibnamefont{and}
  \bibinfo{author}{\bibfnamefont{S.}~\bibnamefont{Franz}},
  \bibinfo{journal}{Europhys. Lett.} \textbf{\bibinfo{volume}{87}},
  \bibinfo{pages}{28001} (\bibinfo{year}{2009}).

\bibitem[{\citenamefont{Vallade and Houchmandzadeh}(2003)}]{VH03}
\bibinfo{author}{\bibfnamefont{M.}~\bibnamefont{Vallade}} \bibnamefont{and}
  \bibinfo{author}{\bibfnamefont{B.}~\bibnamefont{Houchmandzadeh}},
  \bibinfo{journal}{Phys. Rev. E} \textbf{\bibinfo{volume}{68}},
  \bibinfo{pages}{061902} (\bibinfo{year}{2003}).

\bibitem[{XVB()}]{XVBH+03}
\bibinfo{note}{\bibinfo{author}{\bibfnamefont{I.}~\bibnamefont{Volkov}}, \etal,
  \bibinfo{journal}{Nature} \textbf{\bibinfo{volume}{424}},
  \bibinfo{pages}{1035} (\bibinfo{year}{2003}).}

\bibitem[{\citenamefont{Mckane et~al.}(2004)\citenamefont{Mckane, Alonso, and
  {Sol\'e}}}]{MAS04}
\bibinfo{author}{\bibfnamefont{A.~J.} \bibnamefont{Mckane}},
  \bibinfo{author}{\bibfnamefont{D.}~\bibnamefont{Alonso}}, \bibnamefont{and}
  \bibinfo{author}{\bibfnamefont{R.~V.} \bibnamefont{{Sol\'e}}},
  \bibinfo{journal}{Theor. Pop. Biol.} \textbf{\bibinfo{volume}{65}},
  \bibinfo{pages}{67} (\bibinfo{year}{2004}).

\bibitem[{\citenamefont{He}(2005)}]{He05}
\bibinfo{author}{\bibfnamefont{F.}~\bibnamefont{He}},
  \bibinfo{journal}{Functional Ecol.} \textbf{\bibinfo{volume}{19}},
  \bibinfo{pages}{187} (\bibinfo{year}{2005}).

\bibitem[{\citenamefont{Etienne and Alonso}(2006)}]{EA06}
\bibinfo{author}{\bibfnamefont{R.~S.} \bibnamefont{Etienne}} \bibnamefont{and}
  \bibinfo{author}{\bibfnamefont{D.}~\bibnamefont{Alonso}},
  \bibinfo{journal}{J. Stat. Phys.} \textbf{\bibinfo{volume}{128}},
  \bibinfo{pages}{485} (\bibinfo{year}{2006}).

\bibitem[{\citenamefont{Vallade and Houchmandzadeh}(2006)}]{VH06}
\bibinfo{author}{\bibfnamefont{M.}~\bibnamefont{Vallade}} \bibnamefont{and}
  \bibinfo{author}{\bibfnamefont{B.}~\bibnamefont{Houchmandzadeh}},
  \bibinfo{journal}{Phys. Rev. E} \textbf{\bibinfo{volume}{74}},
  \bibinfo{pages}{051914} (\bibinfo{year}{2006}).

\bibitem[{\citenamefont{Babak}(2006)}]{Bab06}
\bibinfo{author}{\bibfnamefont{P.}~\bibnamefont{Babak}},
  \bibinfo{journal}{Phys. Rev. E} \textbf{\bibinfo{volume}{74}},
  \bibinfo{pages}{021902} (\bibinfo{year}{2006}).

\bibitem[{\citenamefont{Chave and Leigh}(2002)}]{CL02}
\bibinfo{author}{\bibfnamefont{J.}~\bibnamefont{Chave}} \bibnamefont{and}
  \bibinfo{author}{\bibfnamefont{E.~G.} \bibnamefont{Leigh},
  \bibfnamefont{Jr.}}, \bibinfo{journal}{Theor. Pop. Biol.}
  \textbf{\bibinfo{volume}{62}}, \bibinfo{pages}{153} (\bibinfo{year}{2002}).

\bibitem[{XCP()}]{XCPL+02}
\bibinfo{note}{\bibinfo{author}{\bibfnamefont{E.}~\bibnamefont{Condit}}, \etal,
  \bibinfo{journal}{Science} \textbf{\bibinfo{volume}{295}},
  \bibinfo{pages}{666} (\bibinfo{year}{2002}).}

\bibitem[{XTi()}]{XTip79}
\bibinfo{note}{The probability distribution function $p(\omega)$ has also been
  advocated for the non-parametric Kolmogorov-Smirnov test to compare taxon
  abundance distributions; see \bibinfo{author}{\bibfnamefont{J.~C.}
  \bibnamefont{Tipper}}, \bibinfo{journal}{Paleobiology}
  \textbf{\bibinfo{volume}{5}}, \bibinfo{pages}{423} (\bibinfo{year}{1979}).}

\bibitem[{\citenamefont{Simpson}(1949)}]{Sim49}
\bibinfo{author}{\bibfnamefont{E.~H.} \bibnamefont{Simpson}},
  \bibinfo{journal}{Nature} \textbf{\bibinfo{volume}{163}},
  \bibinfo{pages}{688} (\bibinfo{year}{1949}).

\bibitem[{XAP()}]{XAPB+06}
\bibinfo{note}{\bibinfo{author}{\bibfnamefont{S.}~\bibnamefont{Azaele}}, \etal,
  \bibinfo{journal}{Nature} \textbf{\bibinfo{volume}{444}},
  \bibinfo{pages}{926} (\bibinfo{year}{2006}).}

\bibitem[{mat()}]{matthew}
\bibinfo{note}{Matthew 25:29 (King James Version): ``For unto every one that
  hath shall be given, and he shall have abundance: but from him that hath not
  shall be taken away even that which he hath.''.}

\bibitem[{\citenamefont{He and Hu}(2005)}]{HH05}
\bibinfo{author}{\bibfnamefont{F.}~\bibnamefont{He}} \bibnamefont{and}
  \bibinfo{author}{\bibfnamefont{X.-S.} \bibnamefont{Hu}},
  \bibinfo{journal}{Ecol. Lett.} \textbf{\bibinfo{volume}{8}},
  \bibinfo{pages}{386} (\bibinfo{year}{2005}).

\bibitem[{\citenamefont{Alonso and McKane}(2004)}]{AM04}
\bibinfo{author}{\bibfnamefont{D.}~\bibnamefont{Alonso}} \bibnamefont{and}
  \bibinfo{author}{\bibfnamefont{A.~J.} \bibnamefont{McKane}},
  \bibinfo{journal}{Ecol. Lett.} \textbf{\bibinfo{volume}{7}},
  \bibinfo{pages}{901} (\bibinfo{year}{2004}).

\bibitem[{\citenamefont{Dornelas et~al.}(2006)\citenamefont{Dornelas, Connolly,
  and Hughes}}]{DCH06}
\bibinfo{author}{\bibfnamefont{M.}~\bibnamefont{Dornelas}},
  \bibinfo{author}{\bibfnamefont{S.~R.} \bibnamefont{Connolly}},
  \bibnamefont{and} \bibinfo{author}{\bibfnamefont{T.~P.}
  \bibnamefont{Hughes}}, \bibinfo{journal}{Nature}
  \textbf{\bibinfo{volume}{440}}, \bibinfo{pages}{80} (\bibinfo{year}{2006}).

\end{thebibliography}

\end{document}